\documentclass[12pt,preprint,usenatbib]{emulateapj}

\usepackage{amssymb}
\usepackage{times}
\usepackage{pifont} 
\usepackage{rotating}

\newcommand{\teff}{$T_{\mathrm{eff}}$}
\newcommand{\muhz}{$\mu$Hz}
\newcommand{\numax}{$\nu_{\mathrm{max}}$}
\newcommand{\dnu}{$\Delta\nu$}

\newcommand{\msol}{M$_\odot$}

\newcommand{\kepler}{\textit{Kepler}}

\shorttitle{Oscillations in red giants observed by K2}
\shortauthors{Stello et al.}

\begin{document}

\title{Oscillating red giants observed during Campaign 1 of the Kepler K2
  mission: New prospects for galactic archaeology}

\author{
Dennis~Stello\altaffilmark{1,2}, 
Daniel~Huber\altaffilmark{1,3,2}, 
Sanjib~Sharma\altaffilmark{1},
Jennifer~Johnson\altaffilmark{4},
Mikkel~N.~Lund\altaffilmark{2}, 
Rasmus~Handberg\altaffilmark{2}, 
Derek~L.~Buzasi\altaffilmark{5}, 
Victor~Silva~Aguirre\altaffilmark{2},
William~J.~Chaplin\altaffilmark{6,2},
Andrea~Miglio\altaffilmark{6,2},
Marc~Pinsonneault\altaffilmark{4},
Sarbani~Basu\altaffilmark{7},
Tim~R.~Bedding\altaffilmark{1,2},
Joss~Bland-Hawthorn\altaffilmark{1},
Luca~Casagrande\altaffilmark{8},
Guy~Davies\altaffilmark{6,2},
Yvonne~Elsworth\altaffilmark{6,2},
Rafael~A.~Garcia\altaffilmark{9},
Savita~Mathur\altaffilmark{10},
Maria~Pia~Di~Mauro\altaffilmark{11},
Benoit~Mosser\altaffilmark{12},
Donald~P.~Schneider\altaffilmark{13,14}, 
Aldo~Serenelli\altaffilmark{15}, and
Marica~Valentini\altaffilmark{16}
}
\altaffiltext{1}{Sydney Institute for Astronomy (SIfA), School of Physics, University of Sydney, NSW 2006, Australia}
\altaffiltext{2}{Stellar Astrophysics Centre, Department of Physics and Astronomy, Aarhus University, Ny Munkegade 120, DK-8000 Aarhus C, Denmark}
\altaffiltext{3}{SETI Institute, 189 Bernardo Avenue, Mountain View, CA 94043, USA}
\altaffiltext{4}{Department of Astronomy, The Ohio State University, Columbus, OH 43210, USA}
\altaffiltext{5}{Department of Chemistry and Physics, Florida Gulf Coast University, Fort Myers, FL 33965, USA}
\altaffiltext{6}{School of Physics \& Astronomy, University of Birmingham, Edgbaston, Birmingham, B15 2TT, UK}
\altaffiltext{7}{Department of Astronomy, Yale University, P.O. Box 208101, New Haven, CT 06520-8101}
\altaffiltext{8}{Research School of Astronomy \& Astrophysics, Mount Stromlo Observatory, The Australian National University, ACT 2611, Australia}
\altaffiltext{9}{Laboratoire AIM, CEA/DSM -- CNRS - Univ. Paris Diderot -- IRFU/SAp, Centre de Saclay, 91191 Gif-sur-Yvette Cedex, France}
\altaffiltext{10}{Space Science Institute,  4750 Walnut street Suite 205 Boulder, CO 80301 USA}
\altaffiltext{11}{INAF, IAPS Istituto di Astrofisica e Planetologia Spaziali, Roma, Italy}
\altaffiltext{12}{LESIA, Observatoire de Paris, PSL Research University, CNRS, Universit\'e Pierre et Marie Curie, Universit\'e Paris
Diderot, 92195 Meudon, France cedex, France}
\altaffiltext{13}{Department of Astronomy and Astrophysics, The Pennsylvania State University, University Park, PA 16802}
\altaffiltext{14}{Institute for Gravitation and the Cosmos, The Pennsylvania State University, University Park, PA 16802}
\altaffiltext{15}{Instituto de Ciencias del Espacio (ICE-CSIC/IEEC) Campus UAB, Carrer de Can Magrans, s/n 08193 Cerdanyola del Valls}
\altaffiltext{16}{Leibnitz Institute f\"ur Astrophysics (AIP), Potsdam, Germany}

\begin{abstract}
NASA's re-purposed \kepler\ mission -- dubbed K2 -- has brought
new scientific opportunities that were not anticipated for the original
\kepler\ mission. One science goal that makes optimal use of K2's
capabilities, in particular its 360-degree ecliptic field of view, is
galactic archaeology -- the study of the evolution of the Galaxy from the 
fossil stellar record.
The thrust of this research is to exploit high-precision, time-resolved
photometry from K2 in order to detect oscillations in red giant stars.
This asteroseismic information can provide estimates of stellar radius (hence
distance), mass and age of vast numbers of stars across the Galaxy.
Here we present the initial analysis of a subset of red giants,
observed towards the North Galactic Gap, during the mission's first full
science  
campaign.  We investigate the feasibility
of using K2 data for detecting oscillations in red giants that span a range
in apparent magnitude and evolutionary state (hence intrinsic luminosity).
We demonstrate that oscillations are detectable for essentially all cool
giants within the $\log g$ range $\sim 1.9$--3.2.  Our detection is 
complete down to $\mathit{Kp}\sim 14.5$, which results in a seismic sample 
with little or no detection bias. This sample is ideally suited to
stellar population studies that seek to investigate potential 
shortcomings of contemporary Galaxy models.
\end{abstract}

\keywords{stars: fundamental parameters --- stars: oscillations --- stars:
  interiors}


\section{Introduction} 
The study of red giant stars has arguably been one of the greatest success
stories of NASA's \kepler\ mission
\citep[e.g.,~][~and references herein]{GarciaStello15}. 
However, a failure of the second of four momentum wheels ended the 
mission in 2013 because the spacecraft could no longer 
acquire stable pointing towards its original field of view.  
Fortunately, ingenious use of the remaining spacecraft capabilities by NASA
and Ball Aerospace engineers rejuvenated the mission as K2 -- a mission
capable of stable pointing at any field along the ecliptic for up to
approximately three months per pointing \citep{Howell14}. 
In this configuration, the \kepler\ roll angle drifts due to a torque 
applied by solar radiation pressure, but this can be counteracted by thruster
firings every six hours to maintain the spacecraft pointing. 
The K2 mission has enabled a broad range
of new science including stellar clusters \citep{Nardiello15}, planets
around bright cool stars
\citep{Crossfield15,Sanchis-Ojeda15,Vanderburg15,Montet15}, solar system 
objects \citep{Szabo15}, stellar activity \citep{Ramsay15}, eclipsing
binaries \citep{Conroy14}, asteroseismology
\citep{Jeffery14,LundHandberg15,Chaplin15} and, in particular,
asteroseismological studies of the Galaxy.

The potential for asteroseismic investigations of large populations of red
giants  
aimed at Galactic studies was recently demonstrated using data from CoRoT and
\kepler\ \citep{Miglio09,Chaplin11,Miglio13,Casagrande15}.
However, the scope of these early studies was limited for two reasons: the
small  
number of distinct direction in the Galaxy probed by those missions, and the
highly complex (and, at some level, not fully documented) selection function of the
observed red giants (Sharma et al. in preparation). 
With K2's 360-degree coverage of the ecliptic the collated efforts from the K2 observing
campaigns, provide a unique opportunity to probe different regions of the
Galaxy, including the thin and thick disks, the halo, and the bulge, based on a
purpose-built selection approach suitable for population studies. 

In this Letter, we present initial results from the K2
Campaign 1 data. Based on a sample of red giants specifically selected to
study stellar populations on a galactic scale, our aim is to determine if
we can obtain an unbiased sample of stars showing oscillations -- a 
crucial first step if these stars are to be used for galactic archaeology
studies.

\section{Observations and light curve preparation}\label{observations}
We used observations obtained as part of the K2 Galactic Archaeology
Program Campaign 1 (C1 proposal
GO1059)\footnote{http://keplerscience.arc.nasa.gov/K2/index.shtml}.   
For the purpose of this Letter, we focused on a targeted set of red giant
candidates based 
on their spectroscopic $\log g$ values ($\log g < 3.8$) drawn from 
APOGEE \citep{Majewski10}, part of the Sloan
Digital Sky Survey III \citep{Eisenstein11}.  The APOGEE survey
used a wide-field multi-object H-band spectrograph \citep{Wilson10} on the
2.5-meter Sloan Foundation telescope at Apache Point
Observatory \citep{Gunn06}.  The red giants discussed here were
observed as part of the main survey \citep{Zasowski13}.  The spectra
were reduced, wavelength-calibrated, and co-added as described
by \citet{Nidever15}.  The determination of stellar parameters
from the the automated pipeline is described by Garcia-Perez et al, {in
preparation). We used the parameters from Data Release
12 \citep{Alam15}; the calibration and verification of the APOGEE
DR 12 results is described by \citet{Holtzman15}.
Of the 121 red giant candidates that satisfy the above-mentioned $\log g$ cut,  
117 were observed in the K2 campaign. The stars span 2.8 dex in $\log g$
and seven magnitudes in 
apparent magnitude ($9\lesssim \mathit{Kp}\lesssim16$), and hence serve as a
suitable benchmark set to characterize the K2 data
fidelity, including the ability to detect oscillations for different levels
of intrinsic and apparent brightness. 

The photometric time series (light curves) from the raw K2 pixel
data are currently not available from NASA. However, light curves
created by 
\citet{VanderburgJohnson14} (VJ hereafter) are publicly
available\footnote{https://www.cfa.harvard.edu/\~{}avanderb/k2.html} and
we used these data in our study.
Power spectra of the light curves denoted `corrected' by VJ are
shown in Fig.~\ref{raw_hipass_gapfill_spectra} (black curves) for three
stars. 
\begin{figure}
\includegraphics[width=8.8cm]{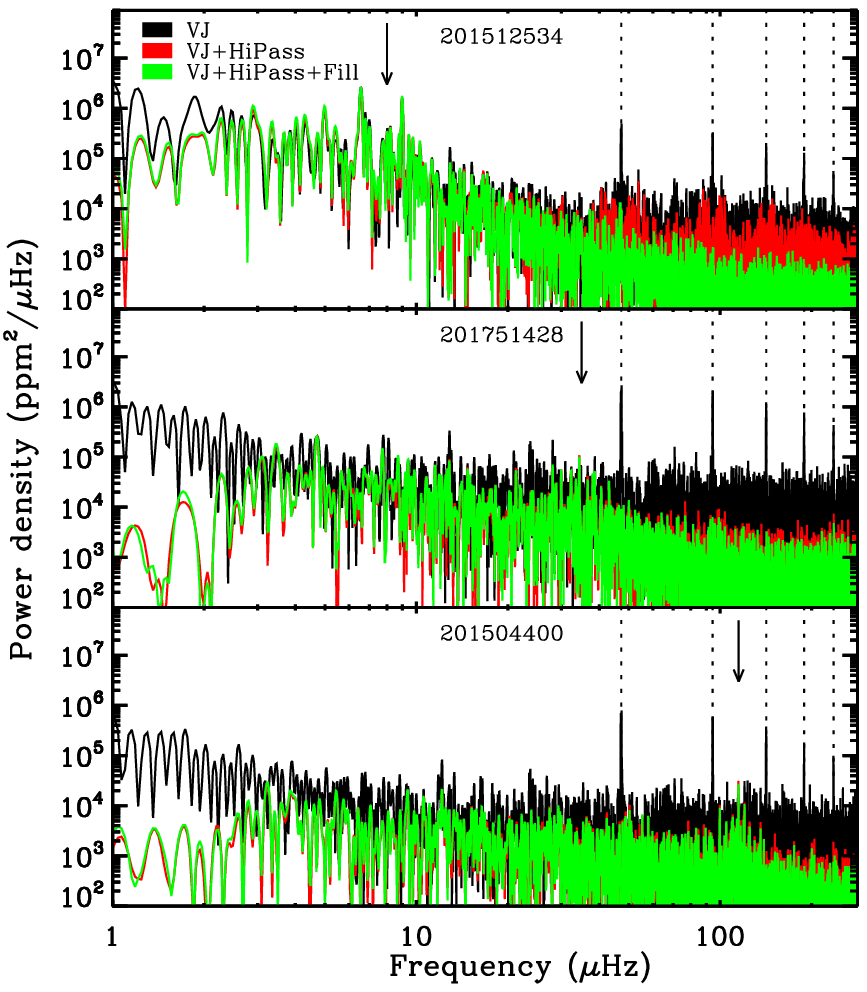}
\caption{Power spectra of three suspected red giants. Their EPIC IDs (Huber et al. 2015, in preparation) are shown in each panel. 
  Arrows mark the location of excess power assumed to be from stellar
  oscillations.  Dotted lines are located at multiples of 47.2281\muhz,
  corresponding to the spacecraft roll angle re-positioning frequency.  Spectra
  in black are based on VJ's `corrected' light curves.  Red spectra show
  the result after high-pass filtering the light curves.  Spectra in green   
  include both a high-pass filter and an interpolation to fill short
  regular gaps in the data caused by the repositioning of the spacecraft.
\label{raw_hipass_gapfill_spectra}} 
\end{figure} 
The 80-day duration of the light curves provides a nominal frequency
resolution of $\sim 0.14\,$\muhz, which imposes a lower limit on
the frequency separation between overtone modes, \dnu, that we can 
reliably determine \citep{Huber10,Hekker12}.  Through the tight correlation
between \dnu\ and \numax\ \citep{Stello09a}, this translates into a
lower limit of about 10\muhz\ in \numax, and hence to the frequency range
in which we can fully characterise the oscillations. 
By restricting our attention to higher-frequency stars, we are then free to
apply a high-pass filter without compromising 
the oscillation signal.  We chose a boxcar filter with a width of two 
days, resulting in a cut-off frequency of about 3\muhz\
(Fig.~\ref{raw_hipass_gapfill_spectra}, red spectra), without
affecting the oscillation signal of our target stars.  
Applying a high-pass filter gives a significant reduction in the
noise floor at all frequencies because it reduces spectral leakage of low-frequency power to higher frequencies, as seen by comparing the black and
the red spectra in Fig.~\ref{raw_hipass_gapfill_spectra}.  

Due to the roughly 6-hour drift and repointing cycle, K2 data generally
show trends on that time scale, and typically one data point is
flagged unsuitable during the repointing \citep[see~][]{VanderburgJohnson14}.
The resulting regular gaps, combined with the
slow drifts, can result in significant leakage of power towards higher
frequencies in the power spectra.  This offset can be avoided if the gaps
are filled \citep{Garcia14,Pires15}, for which we used
linear interpolation 
for all gaps of up to three consecutive data points.  Larger gaps were not filled.  
The interpolation used only the two points that bracketed each gap.
In agreement with 
\citet{Garcia14,Pires15}, we found that the stellar signal was not markedly
affected. However, our ability to detect the oscillations was increased,
particularly for 
stars oscillating at relatively high frequencies.  This improvement is illustrated by
comparing the red and green spectra in
Fig.~\ref{raw_hipass_gapfill_spectra}.   
While gap filling should be used with caution,
it is generally safe when the gaps are short and few compared to the
total number of data points in the light curve
\citep{Garcia14,Pires15}. Here, we filled 3--4\% of all data points.
\begin{figure}
\includegraphics[width=8.8cm]{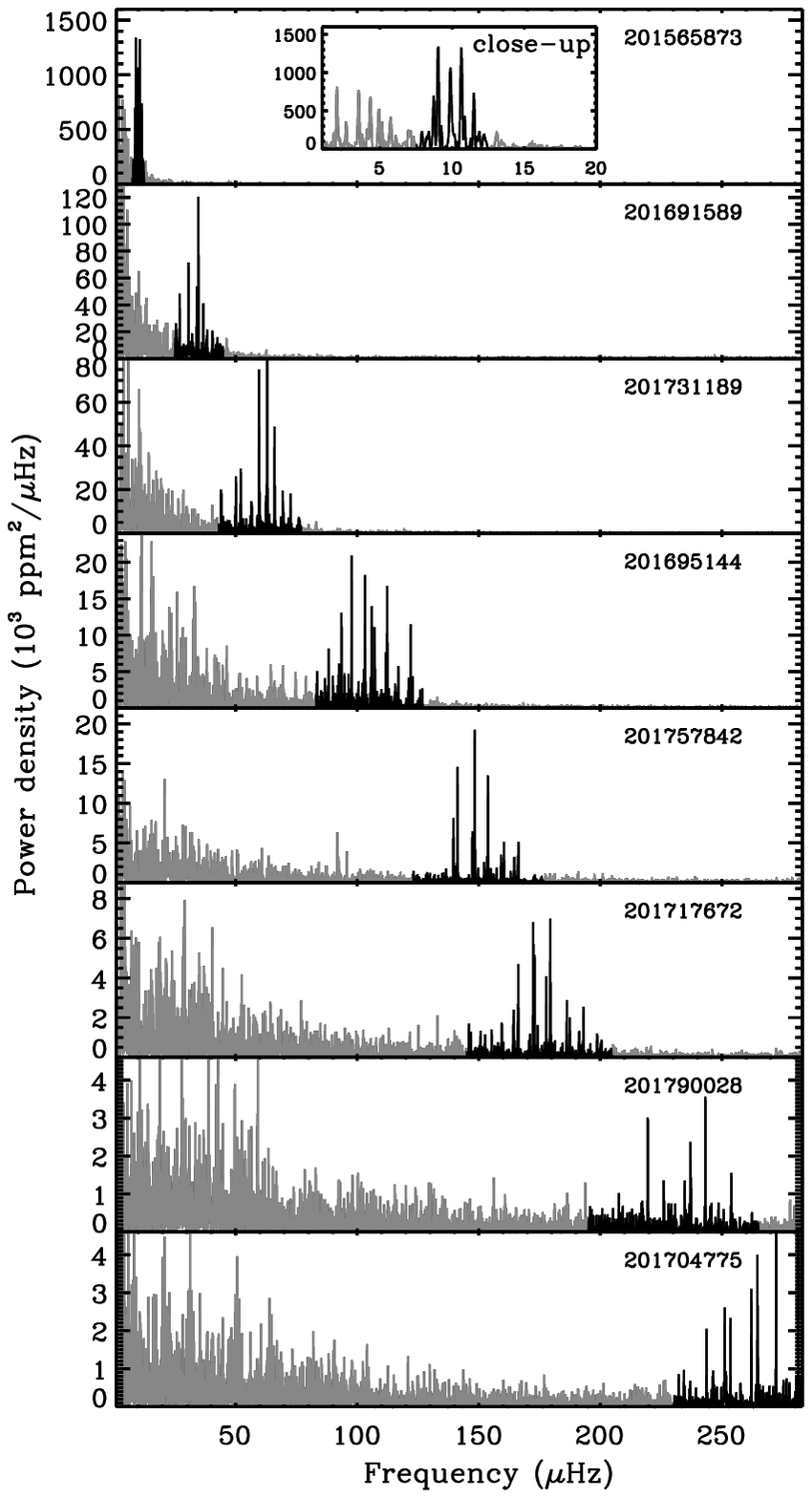}
\caption{Power spectra of selected red giants representing the range in
  \numax, detectable from K2 long-cadence data. Their EPIC ID's are indicated. The stars are ordered from
  most luminous (lowest oscillation frequencies of about 10\muhz) at the
  top, to the least luminous (highest oscillation frequency of about
  270\muhz) at the bottom. The frequency range dominated by the
  oscillations is shown in black.  
\label{example_spectra}} 
\end{figure} 
For these K2 data, the gap filling results in almost complete
removal of peaks in the power spectra at the re-positioning frequency and its
harmonics (Fig.~\ref{raw_hipass_gapfill_spectra}, dotted lines), which are otherwise disruptive for the automated
detection of oscillations and extraction of the global seismic properties.  In the following
analysis, we use the high-pass filtered, gap-filled light
curves. 

\begin{figure*}
\includegraphics[width=18.0cm]{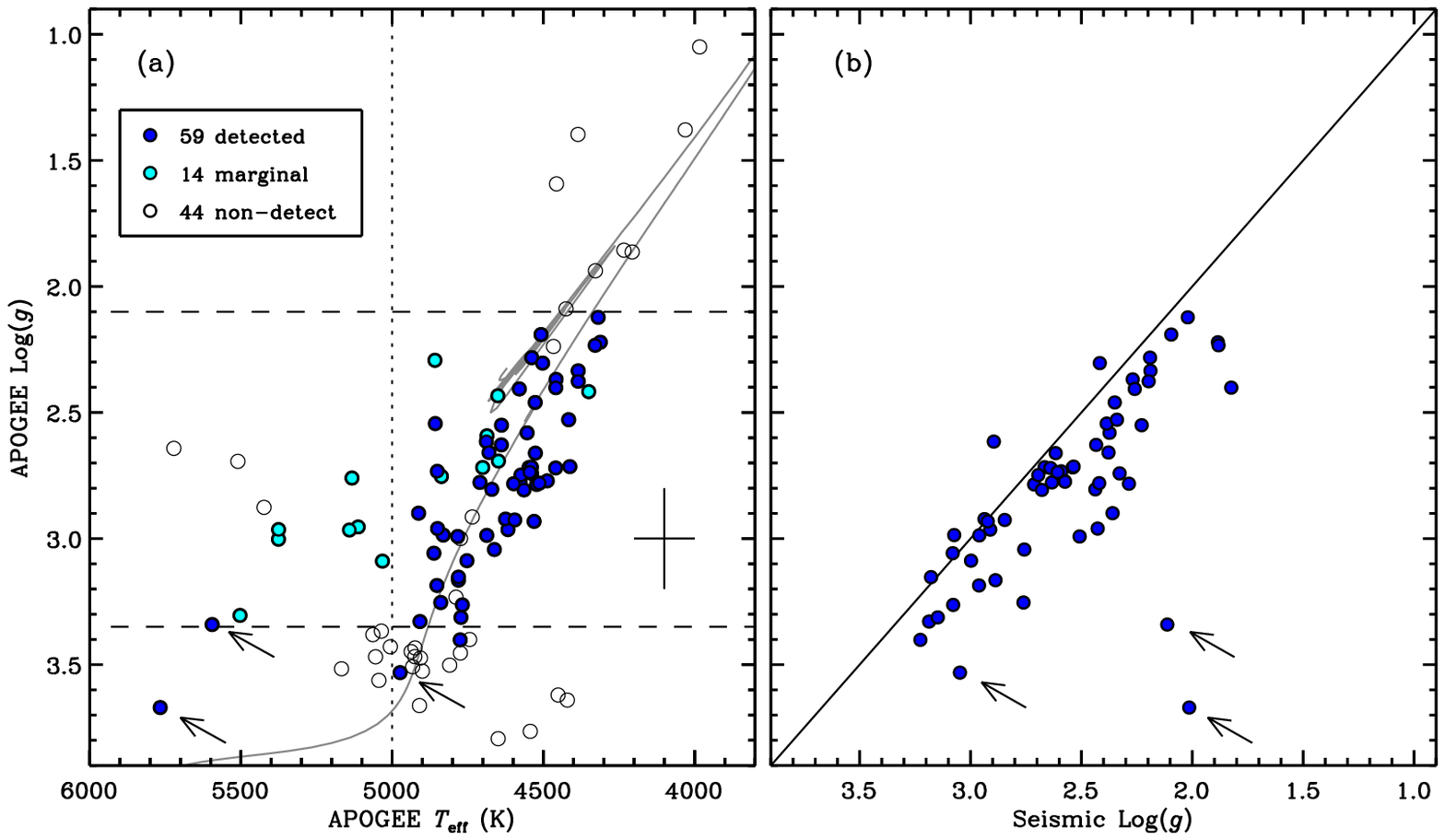}
\caption{(a) $\log g$ versus \teff\ from APOGEE \citep{Majewski10} for all
  117 observed stars in our sample. We plot the values called `raw' in
  Equations 2 and 3 of \cite{Holtzman15} because their `corr'(ected) values
  are not available for all stars in our sample. 
  Filled blue circles mark stars with detected
  oscillations, filled cyan circles show marginal detections, and
  empty circles are stars with no detected oscillations. 
  A typical error bar is shown at ($\log g$,\teff)=(3.0,$4100\,$K). 
  Dashed lines mark the
  $\log g$ range of stars with \numax\ between 10\muhz\ and 270\muhz\ -- the
  detectable range using K2 long cadence data spanning one full campaign
  ($\sim80\,$days). 
  The vertical dotted line marks the upper limit in \teff\ typically found
  for oscillating red giants in previous \kepler\ data
  \citep{Pinsonneault14}. 
  The gray solid curve shows a representative MESA stellar evolution track
  of a 1.2\msol\ model. Arrows indicate stars that potentially have a $\log
  g$ very different from the APOGEE value.
  (b) Relation between APOGEE and seismic $\log g$ of the stars with clear
  seismic detection. The solid line shows a 1-to-1 relationship.
\label{apologg_vs_teff}} 
\end{figure*} 
In future work, we will explore other schemes for directly modelling
instrumental variability. In particular, \citet{Angus15} have recently 
proposed a scheme that alleviates the need for gap filling in time 
series data to produce K2 power spectra that are less sensitive to
systematic effects.

\section{Oscillation analysis}
The power spectra of many stars in our sample reveal clear
oscillations ranging from low-luminosity giants near the bottom
of the red giant branch to stars more luminous than the red clump.  The
detected oscillations cover the  
same frequency range as early \kepler\ light curves of similar length
\citep{Huber10}.   Fig.~\ref{example_spectra} presents power spectra of a
representative subset of our sample.  In each star, we
see the oscillation power excess forming a near regular series of
peaks from overtone modes within a broad envelope, from which \numax\ and
\dnu\ can be measured.  


We carried out a systematic search for oscillations using the
pipeline developed by \citet{Huber09}.  
After inspecting the power spectra and the diagnostic output from
the automated detection algorithm, we classified the stars into three groups. 
A total of 59 stars provided clear detections of both \numax\ and \dnu, 14
stars were classified marginal, and 44 were non-detections.  
Marginal detections refer to stars where either \numax\ or \dnu\ were not determined unambiguously.  Non-detections are mostly those stars
where we did not find any evidence of oscillations. Some were slowly
pulsating (very luminous) giants, which in some cases did show
evidence of oscillation power, but due to the small \numax\ and \dnu\, we
are not confident in claiming those as marginal detections. 
Based on previous experience with \kepler\ data,
most of the marginal detections and almost all of the non-detections
extend over regions in $\log g$-\teff\ space that render them unlikely
to result in measurable \numax\ and \dnu\ values.

To illustrate this property, we show $\log g$ and \teff\
from APOGEE in Fig.~\ref{apologg_vs_teff}a of all 117 stars in our sample
superimposed on a stellar evolutionary track (gray curve) of a 1.2\msol,
roughly solar metallicity MESA model \citep{Paxton11,Paxton13} taken
from \citet{Stello13}.   
From stellar evolution models, we would generally not
expect to find giants hotter than \teff\ $=5000\,$K (except for rare extremely
metal-poor and/or massive stars. Indeed,
previous results from long-term observations by \kepler, showed only two
oscillating giants hotter than APOGEE \teff\ $=5000\,$K
\citep{Pinsonneault14}; one on the lower red giant branch and one
red clump star out of 1916 stars in total.  Our K2 results appear to be in line with those
results, and we should therefore discount stars that appear to be hotter 
than this \teff\ threshold when assessing our detectability capabilities
(Fig.~\ref{apologg_vs_teff}a, vertical dotted line).  
However, we do not advice to apply a \teff\
  selection for future large scale population studies, and only do so
  here to remove what appears as somewhat incompatible \teff\ measurements
  of a few stars in our sample. 
As mentioned in Section~\ref{observations}, oscillations of stars with
\numax\ $\lesssim 10\,$\muhz\ cannot be reliably characterized with 80-day
time series -- the typical length of a K2 campaign.
From the \numax\ $\propto g/$\teff$^{1/2}$ relation \citep{Brown91}, this  
essentially translates into a lower limit on $\log g$, as indicated by the
upper horizontal dashed line (Fig.~\ref{apologg_vs_teff}a).
Similarly, the cadence of the data ($\sim 29.4\,$min) results in a Nyquist
frequency of about 283\muhz, which defines an upper limit on \numax\ of
about 270\muhz, and hence on $\log g$ (Fig.~\ref{apologg_vs_teff}a, lower
horizontal dashed line).  Oscillation frequencies above this limit will be too
close to the Nyquist frequency, compromising automated robust measurement
of both \numax\ and \dnu\ \citep{Stello13}.  This includes stars
oscillating beyond the Nyquist frequency.  
Again, our K2 results confirm these boundaries, with all but three
detected oscillating stars falling within the APOGEE $\log g$ range
2.1--3.35. 

Comparing the APOGEE $\log g$ with the
asteroseismic values, we see in Fig.~\ref{apologg_vs_teff}b that there is an 
offset of about 0.2 dex between the two, in agreement with the findings of
\citet{Holtzman15}.   
Note that we derived the seismic $\log g$ using the above \numax\ scaling
relation with \teff\ from APOGEE.
We also note a large scatter, and in some
cases a very large deviation (stars indicated by arrows).  The seismic
$\log g$ has a typical internal uncertainty of about 0.03 dex 
for the length of data used here \citep{Huber10}, much smaller than
can be obtained from spectroscopy. 
Hence we attribute the scatter in
Fig.~\ref{apologg_vs_teff}b to the uncertainties in the
spectroscopic determinations.  This suggests an RMS scatter of 0.2 dex
of the spectroscopic $\log g$, in agreement with \citet{Pinsonneault14};
for a few stars a deviation of up to 1.5 dex is seen.   
The latter extreme cases could potentially be blends in the K2 data, where a
more evolved, intrinsically brighter star with lower $\log g$ is detected
seismically, while a less evolved star is the source in the APOGEE
spectra.  If they are not blends, the 
two hottest stars marked by arrows must also be much cooler given their
seismic signal, which indicates they are evolved red giants more
luminous than the red clump.
With these uncertainties in mind, we would expect that $\log g$ and \teff\
in Fig.~\ref{apologg_vs_teff}a for some stars are not necessarily 
representative of their true values.  Hence, even in the scenario where we
have 100\% detection rates, we should find some stars with
detected oscillations outside the $\log g$-\teff\ region of expected
detectable oscillations (Fig.~\ref{apologg_vs_teff}a, middle-left region
bracketed by the dashed and dotted lines), and some 
non-detections inside this region; this indeed appears to be the case.

Finally, we examine our success rate in detecting oscillations which,
based on the above considerations, needs to be evaluated only for the stars
expected to show oscillations.
Within the $\log g$-\teff\ region of expected detectable
oscillations, there are 67 stars, of which 55 are clear detections, 7 are
marginal detections, and 5 are non-detections. 
We note that it is not
inconceivable that all 5 non-detections have true values of $\log g$ and
\teff\ that would make them fall outside the detectable region given that
we find a similar number of detections outside the detectable region.
However, some of these stars could be genuinely non-oscillating red giants suppressed by
strong binary interactions \citep{Derekas11,Gaulme14}.
\begin{figure}
\includegraphics[width=8.8cm]{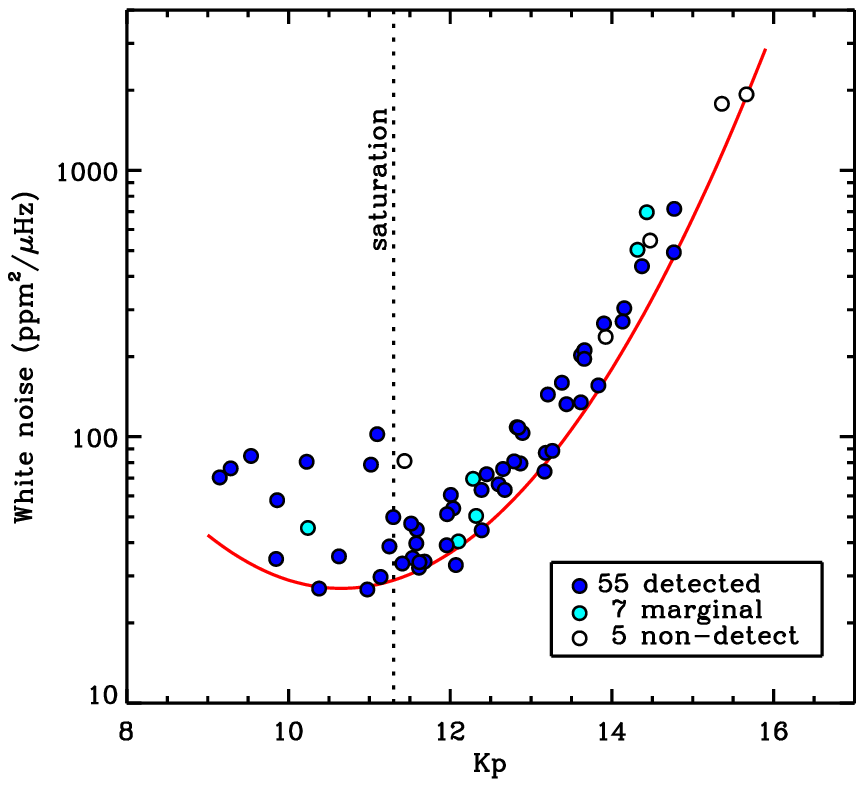}
\caption{Noise level in power spectra as function of {\it Kp} magnitude for
  all stars with 
  APOGEE \teff\ $< 5000\,$K and APOGEE $\log g$ in the range 2.1--3.35
  (see Fig.~\ref{apologg_vs_teff}a).
  The noise is measured as the median power in the range 260-280\muhz.  For
  reference we show a fiducial line described by
  $\log(\mathrm{WhiteNoise})=9.77-1.57\mathit{Kp}+0.074\mathit{Kp}^2$
  ppm$^2$/\muhz\ that follow the lower envelope of the
  magnitude-dependent noise floor (red curve).
  The saturation limit of K2 is indicated by the dotted line.  
  The increased noise for saturated stars can be mitigated using larger
  aperture masks than in the VJ photometry used here
  \citep{LundHandberg15}. 
\label{noise_vs_kpmag}} 
\end{figure} 

To judge whether the non-detections are simply caused by poor photometry, we
show the white noise level versus apparent brightness for all 67 stars in
Fig.~\ref{noise_vs_kpmag}. 
We see no clear trend between the detection category and the 
apparent magnitude, and hence noise level, which suggests that the
non-detections are not generally caused by noisy photometry.  We do see a
possible hint of a faint limit at 
{\it Kp}$\;\gtrsim\;$15 mag, but the low number of faint stars prevents any
definitive conclusion. 
However, for the stars with detected oscillations, we calculated the
correlation 
between {\it Kp} and seismic $\log g$ and found it to be essentially 
zero ($\rho_{\mathit{Kp},\log g}=0.04$).
Hence, even our least-evolved giants (highest $\log g$), 
which have the smallest oscillation amplitudes, span the entire
magnitude range down to $\mathit{Kp}=14.75$.  Taking the uncertainty in
$\mathit{Kp}$ into account and being conservative,
this result suggests that we will be able to detect oscillations in any red giant
down to $\mathit{Kp} \sim 14.5$, as long as they are within the required
$\log g$ range, which for the seismic $\log g$ is $\sim$ 1.9 -- 3.2.

\section{Summary and outlook}
We have performed initial asteroseismic analyzes of K2 C1 data for
over one hundred stars expected to be red giants based on their
spectroscopically determined $\log g$ and \teff.  
We detect oscillations in almost all stars cooler than
$5000\,$K within 2.1--3.35 in $\log g$ \citep[on the scale of the `raw' data in~][]{Holtzman15}, which 
comprise the target stars of ongoing K2-based galactic archaeology studies. 
The results indicate that our detection rate is complete
down to $\mathit{Kp} \sim 14.5$, represented by the faintest oscillating
star in our sample ($\mathit{Kp} = 14.75$) -- a low-luminosity and hence
low-intrinsic-amplitude giant with a detected \numax\ of 150\muhz. 
More stars fainter than $\mathit{Kp} = 14.5$ need to be analyzed in order to
determine the true faint limit, which will set the limit for a magnitude-complete sample of red giants within the 1.9--3.2 range in seismic
$\log g$.  Knowing this selection function is crucial for galactic archaeology.

We note that the characteristic timescale for the spacecraft attitude control
system to react to pointing errors was adjusted from C3 onwards to obtain
lower pointing jitter (by about a factor of 3-4), which is expected to result in a
total noise level only 30\% higher than in the original \kepler\ mission
(Doug Caldwell, private communication).  We anticipate that this would
push the faint end of the detection limit to even fainter stars than the C1
results reported here.

Looking ahead, the prospects for characterizing oscillations in red giants using
K2 data are very promising.  We can expect to be magnitude
complete in our detection rates down to $\mathit{Kp} \sim 14.5$, and
possibly fainter in future campaigns, which is strong affirmation for using red 
giants as distant probes of
the Galaxy's structure and evolution out to at least $\sim 7\,$kpc.  Of
order 5000--10000 targets have already been observed during each of the first five K2
observing campaigns, 
and similar numbers are expected to follow throughout its mission,
providing several tens of thousands of red giants with detected oscillations.
This sample will grow even further with the launch of NASA's next
all-sky planet-finding mission, TESS \citep{Ricker14}, and on the time
scale of a decade we can expect significant boosts in sample size from
missions like WFIRST \citep{Gould15WFIRST}, Euclid \citep{Gould15EUCLID},
and PLATO \citep{Rauer14}.

\acknowledgments
This work was supported by the National Science Foundation
under grants PHY 11-25915 and AST 11-09174.
Funding for the Stellar Astrophysics Centre is provided by The Danish
National Research Foundation (Grant DNRF106). The research is supported by
the ASTERISK project (ASTERoseismic Investigations with SONG and Kepler)
funded by the European Research Council (Grant agreement no.: 267864). 
D.S. acknowledges support from the Australian Research Council.
D.H. acknowledges support by the Australian Research Council's Discovery
Projects funding scheme (project number DE140101364) and support by the
National Aeronautics and Space Administration under Grant NNX14AB92G issued
through the Kepler Participating Scientist Program. 
Funding for SDSS-III has been provided by the Alfred P. Sloan Foundation,
the Participating Institutions, the National Science Foundation, and the
U.S. Department of Energy Office of Science. The SDSS-III web site is
http://www.sdss3.org/. 
SDSS-III is managed by the Astrophysical Research Consortium for the
Participating Institutions of the SDSS-III Collaboration including the
University of Arizona, the Brazilian Participation Group, Brookhaven
National Laboratory, Carnegie Mellon University, University of Florida, the
French Participation Group, the German Participation Group, Harvard
University, the Instituto de Astrofisica de Canarias, the Michigan
State/Notre Dame/JINA Participation Group, Johns Hopkins University,
Lawrence Berkeley National Laboratory, Max Planck Institute for
Astrophysics, Max Planck Institute for Extraterrestrial Physics, New Mexico
State University, New York University, Ohio State University, Pennsylvania
State University, University of Portsmouth, Princeton University, the
Spanish Participation Group, University of Tokyo, University of Utah,
Vanderbilt University, University of Virginia, University of Washington,
and Yale University.  


\end{document}